\documentclass[twocolumn,showpacs,amsmath,preprintnumbers,aps,prb,superscriptaddress]{revtex4-1}

\usepackage{graphicx} 
\usepackage[english]{babel}  
\usepackage[latin1]{inputenc}
\usepackage{bm}
\usepackage{wrapfig}
\usepackage{SIunits}
\usepackage{longtable}

\begin{document}

\title{The EDMR Microscope -- Combining Conductive Atomic Force Microscopy with Electrically Detected Magnetic Resonance}  

\author{Konrad Klein}
\email[Corresponding author: ]{konrad.klein@wsi.tum.de}
\affiliation{Walter Schottky Institut, Technische Universit\"{a}t M\"{u}nchen, Am Coulombwall 4, 85748 Garching, Germany}
\author{Benedikt Hauer}
\altaffiliation[Present address: ]{I. Physikalisches Institut, RWTH Aachen, 52056 Aachen, Germany}
\affiliation{Walter Schottky Institut, Technische Universit\"{a}t M\"{u}nchen, Am Coulombwall 4, 85748 Garching, Germany}
\author{Benedikt Stoib}
\affiliation{Walter Schottky Institut, Technische Universit\"{a}t M\"{u}nchen, Am Coulombwall 4, 85748 Garching, Germany}
\author{Markus Trautwein}
\affiliation{Walter Schottky Institut, Technische Universit\"{a}t M\"{u}nchen, Am Coulombwall 4, 85748 Garching, Germany}
\author{Sonja Matich}
\affiliation{Walter Schottky Institut, Technische Universit\"{a}t M\"{u}nchen, Am Coulombwall 4, 85748 Garching, Germany}
\author{Hans Huebl}
\affiliation{Walther-Meißner-Institut, Bayerische Akademie der Wissenschaften, Walther-Meißner-Str.~8, 85748 Garching, Germany}
\author{Oleksandr Astakhov}
\affiliation{IEK5-Photovoltaik, Forschungszentrum Jülich, Leo-Brandt-Straße, 52425 Jülich, Germany}
\author{Friedhelm Finger}
\affiliation{IEK5-Photovoltaik, Forschungszentrum Jülich, Leo-Brandt-Straße, 52425 Jülich, Germany}
\author{Robert Bittl}
\affiliation{Freie Universität Berlin, Fachbereich Physik, Arnimallee 14, 14195 Berlin, Germany}
\author{Martin Stutzmann}
\affiliation{Walter Schottky Institut, Technische Universit\"{a}t M\"{u}nchen, Am Coulombwall 4, 85748 Garching, Germany}
\author{Martin S. Brandt}
\affiliation{Walter Schottky Institut, Technische Universit\"{a}t M\"{u}nchen, Am Coulombwall 4, 85748 Garching, Germany}

\begin{abstract}
\noindent We present the design and implementation of a scanning probe microscope, which combines electrically detected magnetic resonance (EDMR) and (photo-)conductive atomic force microscopy ((p)cAFM). The integration of a 3-loop 2-gap X-band microwave resonator into an AFM allows the use of conductive AFM tips as a movable contact for EDMR experiments. The optical readout of the AFM cantilever is based on an infrared laser to avoid disturbances of current measurements by  absorption of straylight of the detection laser. Using amorphous silicon thin film samples with varying defect densities, the capability to detect a spatial EDMR contrast is demonstrated. Resonant current changes as low as \unit{20}{fA} can be detected, allowing the method to realize a spin sensitivity of $8 \times \unit{10^6}{spins/\sqrt{Hz}}$ at room temperature.\\
\end{abstract}

\maketitle

\section{Introduction}
\label{introduction}

Electrically detected magnetic resonance (EDMR) has proven to be a powerful technique to understand the influence of paramagnetic states on electronic transport and recombination properties of semiconductors. It was successfully applied to a wide variety of material systems, ranging from inorganic to organic semiconductors and to devices such as thin film solar cells or transistors.\cite{Schmidt1966,Stutzmann2000,Lips2003,Pereira2009,Behrends2010,Baker2011,Lo2012} One of the major advantages of EDMR compared to conventional electron spin resonance (ESR) is its higher sensitivity. A systematic study of the sensitivity of continuous wave EDMR (cwEDMR) at liquid helium temperature using phosphorus-doped silicon samples has shown that at present cwEDMR is able to detect as few as 100 spins.\cite{McCamey2006a}  Still, the active area of most EDMR samples is in the range of $\rm{mm}^2$ corresponding to approximately $10^8$ spins for low spin density samples (assuming an active area of $\unit{1}{mm}^2$, a sample thickness of $\unit{500}{nm}$ and $\unit{10^{15}}{cm^{-3}}$ as the density of paramagnetic states typical for amorphous semiconductors). Already in 1995 Stich~\textit{et al.} pointed out that the sensitivity of cwEDMR should be sufficient for imaging experiments.\cite{Stich1995} They suggested to perform mapping experiments by scanning the sample surface with a laser beam and recording an EDMR spectrum for each point of the scan. In analogy to cwEDMR and pulsed EDMR (pEDMR)\cite{Boehme2003,Stegner2006} this would be called scanning EDMR (sEDMR). The only EDMR imaging experiments published to date were performed by Sato~\textit{et al.} between 2000 and 2004.\cite{Sato2000,Sato2001,Sato2004} They developed a $\unit{900}{MHz}$ EDMR imaging system employing a gradient field technique to obtain EDMR images.\cite{Sato2000,Sato2001} Using this setup the authors could visualize the recombination pattern of photo-generated charge carriers at a $\mbox{Si/SiO}_2$-interface with a spatial resolution better than $\approx 2~\mbox{mm}$.\cite{Sato2001}

In contrast to EDMR, there is a strong ongoing effort to improve the spatial resolution of ESR imaging setups to achieve sub-micrometer resolution. One approach is the work of Blank~\textit{et al.} based on a local microwave excitation of the spin system by surface loop-gap microresonators in cryogenic probe heads to detect as little as several hundred spins in a voxel of $0.5 \times 0.75 \times \unit{5}{\mu m^3} $ at low temperatures.\cite{Blank2013} The probe head designed by these authors is already set up for the electrical detection of magnetic resonance transitions as well.  

In addition to these ``conventional'' detection schemes, various magnetic resonance imaging experiments based on scanning probe techniques were reported. Utilizing magnetic force microscopy (MFM), Rugar~\textit{et al.} developed a magnetic resonance force microscopy (MRFM) setup allowing the detection of a single electron spin.\cite{Rugar2004} By increasing the sensitivity and combining MRFM with 3D image reconstruction they were able to image the $^1$H spin density in tobaco mosaic viruses with a spatial resolution of better than $\unit{10}{nm}$.\cite{Degen2009,Poggio2010} By combining scanning tunneling microscopy (STM) with very sensitive radio frequency (rf) detection, several groups were able to observe paramagnetic species on non-paramagnetic surfaces by detecting a weak resonance in the tunneling current at the Larmor frequency corresponding to the $g$-factor of the paramagnetic species under investigation.\cite{[{For a review see }] Balatsky2012} This approach is called ESR-STM. Another approach is based on the interaction of spins in a sample with the spin of a single NV$^-$ center in diamond read out using an optically detected magnetic resonance (ODMR) scheme. Using this technique the detection of individual spins on a surface has been demonstrated.\cite{Maze2008,Grinolds2011,Maletinsky2012,Grinolds2013}

Single spin detection was also achieved by experiments not based on scanning probe techniques. Using single electron transistors, Elzermann~\textit{et al.}~and~Morello~\textit{et al.} were able to read out the spin state of a single electron spin.\cite{Elzerman2004,Morello2010} Single spin sensitivity in ODMR has been achieved already in 1993 by Köhler~\textit{et al.}~and Wrachtrup~\textit{et al.}.\cite{KOHLER1993, WRACHTRUP1993}
  
In the present publication we report on the combination of another scanning probe technique, namely conductive atomic force microscopy (cAFM) or photo-conductive AFM (pcAFM), with EDMR in an instrument which we name EDMR microscope. (p)cAFM as a non-spin-dependent technique is widely used to probe the topography and the local conductivity of samples simultaneously by scanning across the sample surface with a conductive AFM tip with an applied bias. To mention a few examples, (p)cAFM was used to investigate the local transport properties of amorphous silicon and microcrystalline silicon,\cite{Koida1996,Azulay2005,Ledinsky2011} organic semiconductor devices\cite{Alexeev2006,Coffey2007} or nanostructures.\cite{Ou2010,Hwang2011}

\begin{figure}
	\centering
	\includegraphics[width=0.46\textwidth]{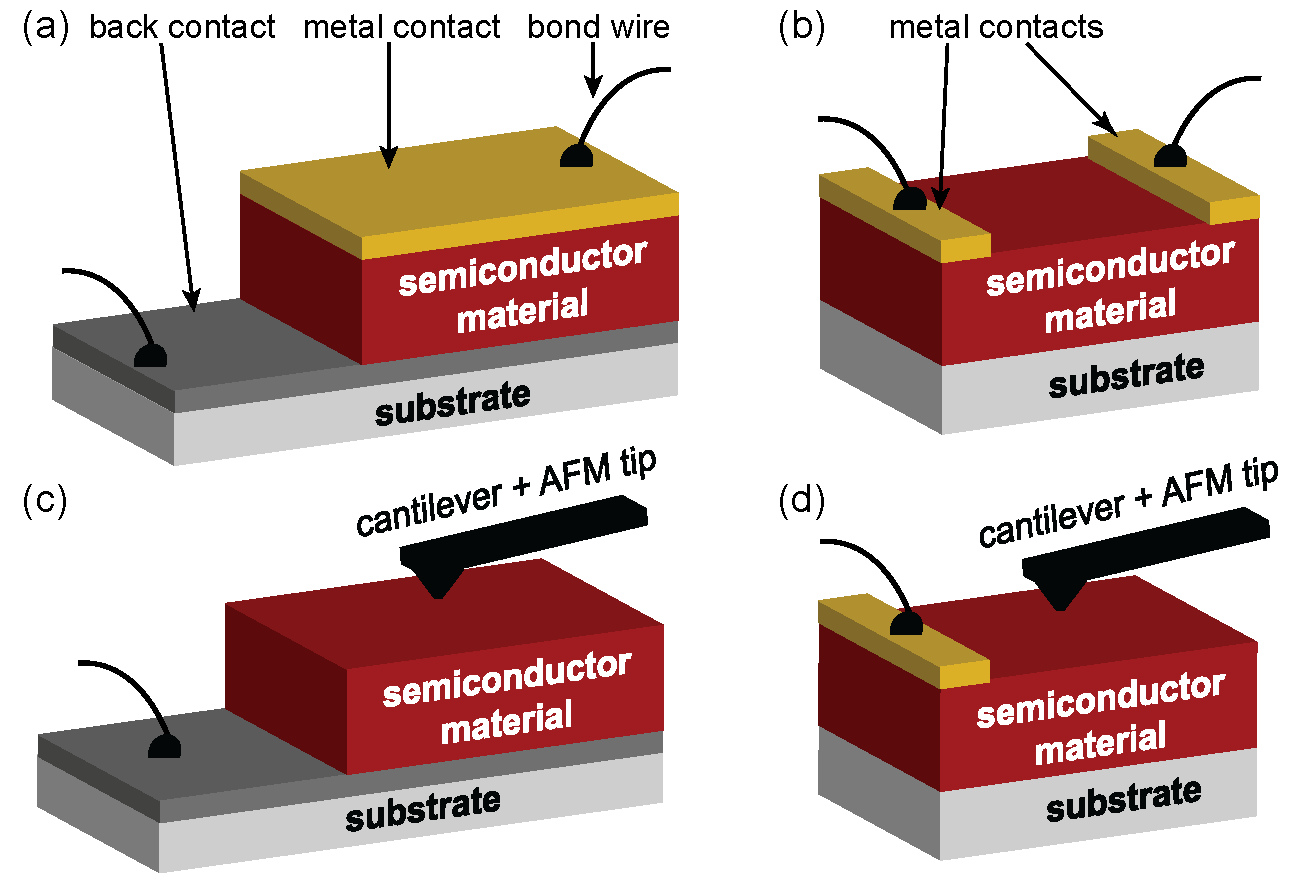}
	\caption{\label{fig1} Panels (a) and (b) schematically show contact geometries frequently used in EDMR experiments. (c) and (d) depict the corresponding contact geometries used for the EDMR microscope.} 
\end{figure}

%
%
By merging (p)cAFM and EDMR we combine the ability to investigate the topography and the local conductivity with the possibility to obtain spectroscopic information on the relevant spin-dependent transport mechanisms in contrast to conventional EDMR probing only global properties. To test the integration of both techniques into one setup, we focus on two contact geometries depicted in Figs.~\ref{fig1}~(c)~and~\ref{fig1}~(d), where one of the typically lithographically defined contacts is replaced by a conductive scanable AFM tip. For comparison, Fig.~\ref{fig1}~(a) shows the sandwich-like contact geometry and Fig.~\ref{fig1}~(b) shows the coplanar contact geometry frequently used in standard EDMR experiments.   

In the following, we briefly introduce the principle of EDMR (Sec.~\ref{EDMRintro}) before we identify relevant design criteria of the EDMR microscope (Sec.~\ref{design}). We introduce the actual design of the microscope in detail in Sec.~\ref{microscopesystem}, and discuss its room temperature operation, performance and sensitivity (Sec.~\ref{operation}), comparing the latter to previous ESR and EDMR experiments in Sec.~\ref{discussion}.

\section{A Short Introduction to EDMR}
\label{EDMRintro}

\begin{figure}
	\centering
	\includegraphics[width=0.46\textwidth]{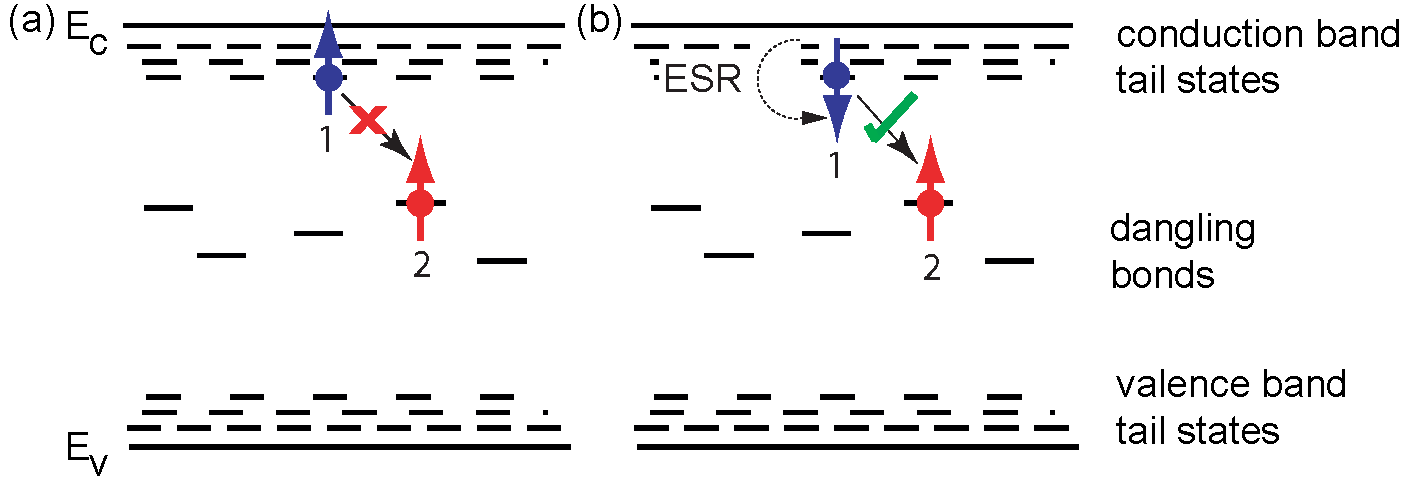}
	\caption{\label{fig2} Schematic representation of a spin-dependent recombination step involving an electron in a conduction band tail state and a paramagnetic dangling bond state in hydrogenated amorphous silicon.} 
\end{figure}

The results presented later in this publication were obtained using hydrogenated amorphous silicon (a-Si:H), a material well established in thin-film photovoltaics. Here, we will focus on the signal formation in electrically detected magnetic resonance due to spin-dependent recombination of the photo-generated charge carriers in a-Si:H at room temperature to point out the working principle of EDMR.
The schematic band structure of undoped a-Si:H is depicted in Fig.~\ref{fig2}. In contrast to crystalline silicon, in a-Si:H conduction and valence band tail states exist due to the disorder in the material.\cite{Street1991} In addition to these tail states, there is a certain density of dangling bond states in the middle of the band gap.\cite{Street1991} If the spin of an electron trapped in a conduction band tail state (blue arrow labeled as state 1)  and the spin of the unpaired electron of a dangling bond in the middle of the band gap (red arrow labeled as state 2) are aligned parallelly, a transition of the electron from state 1 into state 2 is suppressed due to the Pauli principle as shown in Fig.~\ref{fig2}~(a). By driving ESR transitions on one of the two spin states, for example state 1 (Fig.~\ref{fig2}~(b)), the spin of the electron trapped in the tail state is flipped and the relative spin orientation is changed from parallel to antiparallel. As a consequence the recombination of the electron from state 1 into state 2 is possible (Fig.~\ref{fig2}~(b)). 

To induce the ESR transitions in a conventional paramagnetic spin 1/2 system, the sample must be placed in an external magnetic field $B_0$, resulting in a Zeemann splitting of the two states with magnetic quantum numbers $m=\pm 1/2$ by $\Delta E = g \mu_\mathrm{B} B_0$. Here, $g$ is the $g$-factor, which is characteristic for the paramagnetic state under investigation, and $\mu_\mathrm{B}$ the Bohr magneton.\cite{Pake1973} Applying a microwave magnetic field $B_\mathrm{1} \bot B_{0}$ with a frequency $\nu = g \mu_\mathrm{B} B_0/h$ to the sample (with $h$ being Planck's constant), magnetic dipole transitions between the $m=\pm 1/2$ states are excited.\cite{Spaeth2003}

For a-Si:H the transition described above enhances the recombination rate, which is observed experimentally in EDMR as a resonant quenching of the photo-current at $g \approx 2.005$ with a peak-to-peak linewidth $\Delta B_{\mathrm{pp}} \approx \unit{1}{mT}$.\cite{Dersch1983,Stutzmann2000} The signal amplitude observed can be tuned by taking advantage of light-induced degradation of a-Si:H, the so-called Staebler Wronski effect. \cite{Staebler1977, Dersch1981, Hirabayashi1980} If an annealed a-Si:H sample is illuminated with intense (above bandgap) light, the photo-conductivity decreases due to a reversible increase of the density of dangling bonds, which act as recombination centers. This increase of the dangling bond density can also be observed as a higher EDMR signal amplitude compared to the EDMR signal amplitude determined for the initial annealed state.\cite{Street1982} These changes can be fully reverted by re-annealing the sample at $T \approx \unit{450}{K}$ in the dark for about 1 hour.\cite{Staebler1977}

\section{Design Criteria for the EDMR microscope}
\label{design}

An AFM system serves as a starting point for the development of the EDMR microscope. In addition to the usual operating modes to obtain the topography of a sample, this AFM system must be set up for \mbox{(photo-)}conductive AFM measurements. To avoid disturbances of the (p)cAFM operation by the AFM system itself, the detection of the AFM cantilever deflection must not influence the transport measurements. In particular, if an optical detection is used, the wavelength of the detection laser must be chosen such that its straylight is not strongly absorbed by the samples investigated preventing the excitation of photo-generated charge carriers. 

To drive the ESR transitions the sample has to be additionally positioned in an antinode of the microwave magnetic field $B_1$ (see Sec.~\ref{EDMRintro}). This can be achieved by placing the sample inside a resonator, which also separates the magnetic and electric field components of the microwave field. This requires that the resonator is integrated between the sample holder and the AFM head, at which the cAFM cantilever is mounted. In addition, a system for controlled illumination of the sample has to be added, since in the majority of EDMR experiments resonant changes of the photo-conductivity are detected. Typically detected relative resonant current changes $\Delta I / I$ are in the range of $10^{-7}\ldots10^{-3}$, which requires lock-in detection schemes.\cite{Stutzmann2000} The necessary modulation of the signal can be achieved by either microwave frequency or amplitude modulation or by magnetic field modulation.\cite{Lee2012} The latter requires a pair of modulation coils next to the resonator. To characterize the dynamics of the transport processes observed in EDMR as completely as possible, it is necessary to perform experiments at different temperatures, as important transport properties like hopping, trapping or detrapping rates as well as spin properties like the spin-lattice relaxation time $T_1$ can depend significantly on temperature.\cite{Tiedjea,Cullis1975,Stutzmann1983} As a consequence, the operating temperature range of the EDMR microscope should extend from room temperature down to liquid helium temperature.

Finally, to be able to use standard ESR equipment, the EDMR microscope should be compatible to, e.g., existing X-Band ESR/EDMR spectrometers. Therefore, the microwave frequency $\nu$ should be $\nu \approx \unit{9 \ldots 10}{GHz}$ and the EDMR microscope should fit into existing ESR electromagnets, limiting the outer diameter of the EDMR microscope setup including cryostat to $\approx \unit{70}{mm}$, which is the typical distance between the pole shoes of corresponding magnets. For these microwave frequencies the AFM system must withstand magnetic fields $B_0$ of up to \unit{0.4}{T} assuming $g \approx 2$.

\section{The EDMR microscope system}
\label{microscopesystem}

\subsection{Microscope Head}
\label{head}

\begin{figure*}
	\includegraphics[width=0.92\textwidth]{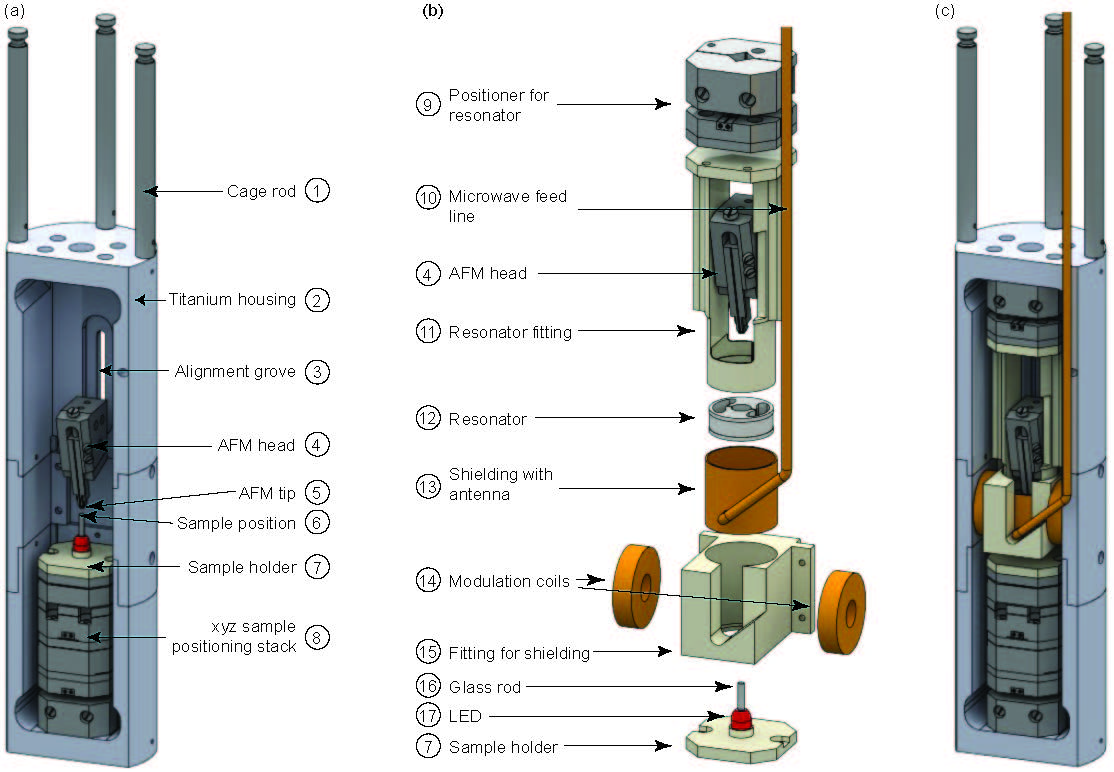}
	\caption{\label{fig3} (a) To-scale model of the microscope head without the microwave assembly. (b) Explosion view of the microwave assembly and the AFM head, the modulation coils and the sample holder. (c) To-scale model of the microscope head with the microwave assembly integrated. For the sake of clarity the electrical wiring of the components, the microscope stick to which the microscope head is attached via the cage rods \textcircled{1} and the fiber to read out the AFM cantilever deflection are not shown.} 
\end{figure*}

The microscope head of the EDMR microscope is a modified commercial attoAFMI microscope head (attocube systems), able to operate in high magnetic fields down to  liquid helium temperature and fulfilling the geometric requirements presented in Sec.~\ref{design}. It is shown as a to-scale CAD-model without the additional EDMR-related components in Fig.~\ref{fig3}~(a). For the sake of clarity the electrical wiring of the components as well as the microscope stick to which the microscope head is attached via the cage rods \textcircled{1} are not shown. All components of the microscope head are placed within a 3-piece housing made of non-magnetic titanium \textcircled{2}. The outer diameter of the housing is $\approx \unit{5}{cm}$ and the dimensions of the hollow inner area of the housing are $\unit{30}{mm} \times \unit{26}{mm}$, limitting the lateral size of any additional EDMR components to $\approx \unit{30}{mm}$. The bottom part of the housing is used as a base for the $xyz$ sample positioning stack \textcircled{8}. This stack combines a coarse positioning unit and a scanner unit. The coarse positioning unit consists of three piezo positioners for the $x$-, $y$-, $z$-directions (2x~ANPx101, 1x~ANPz101), which have a travel range of $\unit{5}{mm}$ each. On top of the coarse positioning unit the scanning unit is placed, which consists of a $xy$-scanner (ANSxy100, scan range at room temperature (RT) $\approx \unit{50}{\mu m}$ in each direction, at low temperature (LT) $\approx \unit{30}{\mu m}$ in each direction) and a $z$-scanner (ANSz100, scan range at RT $\approx \unit{13}{\mu m}$, at LT $\approx \unit{11}{\mu m}$). The sample holder \textcircled{7}, which is described in detail in Sec.~\ref{holder}, is mounted on top of the $z$-scanner.

The upper part of the housing is used to mount the AFM head \textcircled{4}. It is placed in a grove \textcircled{3} in the housing, which helps to align the AFM head and guides the AFM head, when it is slid up or down to be positioned in close proximity to the sample. The AFM head \textcircled{4} itself is an attocube AFM head without the clamping mechanism to mount the AFM cantilever, since the necessary mechanics would block space needed for the additional microwave components. Instead, the AFM cantilever is glued onto the AFM cantilever holder directly. To read out the deflection of the AFM cantilever, attocube's fiber interferometer-based deflection-sensing technique (the fiber is not shown) based on an infrared laser emitting at $\approx \unit{1300}{nm}$ is used, instead of the standard AFM cantilever deflection sensing schemes based on laser diodes emitting visible red light.\footnote{For further details on the AFM head and the interferometric readout scheme, see www.attocube.com} Red light ($\approx \unit{1.9}{eV}$) would be strongly absorbed in most silicon-based semiconductors like crystalline, microcrystalline or amorphous silicon and would lead to a misinterpretation of pcAFM data as discussed in detail by Ledinsk\'{y}~\textit{et al.}.\cite{Ledinsky2011} In contrast, infrared light ($\approx \unit{0.95}{eV}$) is only very weakly absorbed.\cite{Street1991} To be able to measure the local (photo-)conductivity of a sample, an additional electrical connection to contact the AFM tip was integrated in the AFM head (not shown).

\subsection{Microwave Assembly}

In addition to the standard AFM parts the microwave assembly has to be integrated into the microscope head. In particular the microwave resonator has to be positioned between the sample holder \textcircled{7} and the AFM head \textcircled{4} in Fig.~\ref{fig3}~(a). Usual ESR X-band resonators such as rectangular cavities do not fit into the setup, since their side length is of the order of the wavelength, i.e.~$\sim \unit{3}{cm}$ at \unit{9}{GHz}, and therefore only little to no space would be left next to the resonator for additional components like modulation coils.  In addition, the AFM head itself, consisting of electrically conductive material, has to be placed in very close vicinity of the sample and the resonator. This configuration would be geometrically very challenging when using a standard rectangular cavity.

Another approach is to use an X-band loop-gap resonator.\cite{Froncisz1982} Such a resonator is a so-called lumped circuit, with the physical dimensions of the resonator being smaller than the microwave wavelength at the resonant frequency. In addition, loop gap resonators combine high filling factors and a good separation of electric and magnetic field, exhibit reasonably high quality factors $Q$ and are comparably robust against conductive material in their close vicinity.\cite{Froncisz1982} 

\begin{figure}
	\centering
	\includegraphics[width=0.46\textwidth]{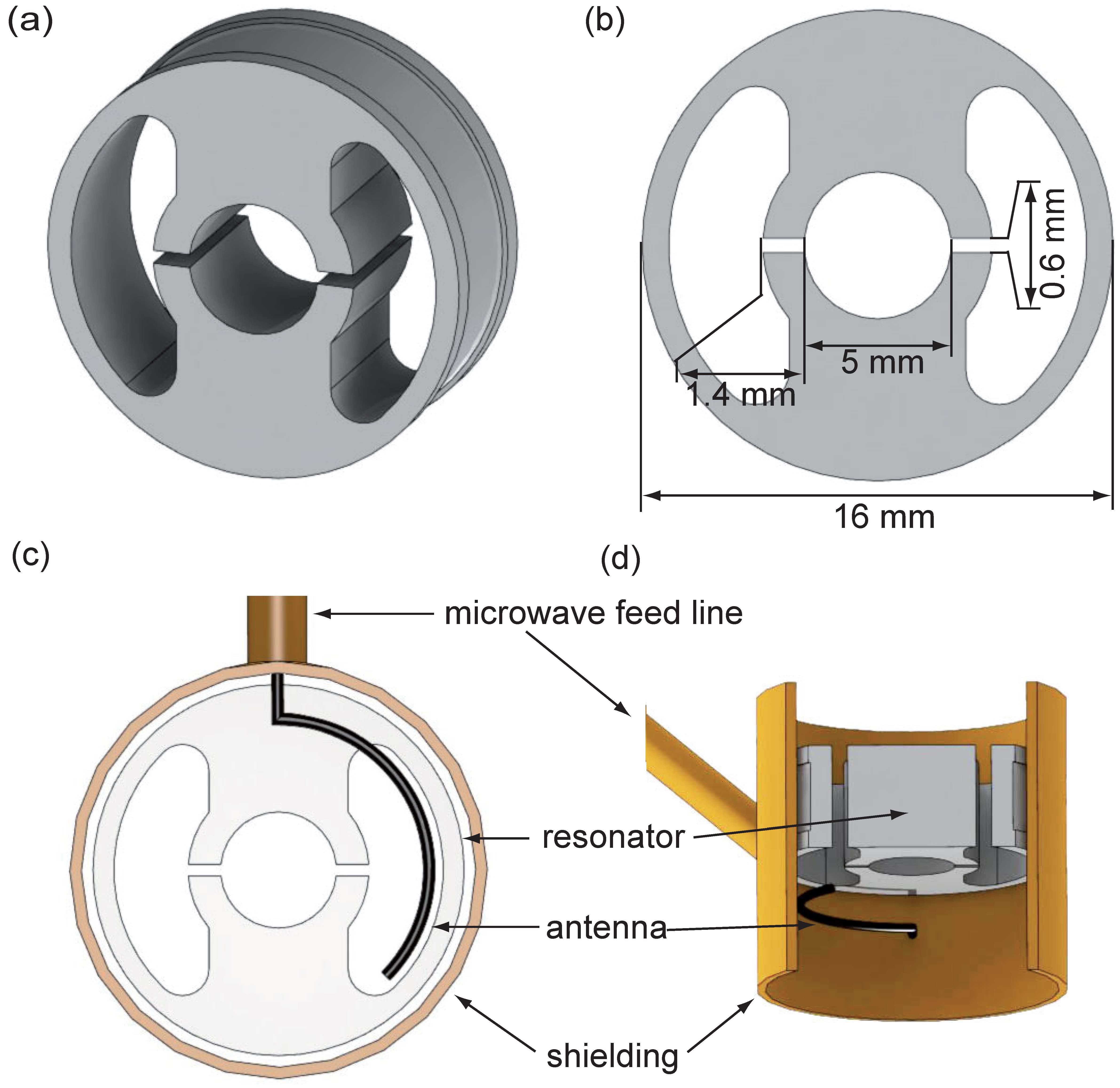}
	\caption{\label{fig4} (a) Model of the 3-loop 2-gap resonator used in this work. (b) To-scale front view of the resonator. (c) Bottom view of the shielding including the resonator to show the orientation of the resonator with respect to the shielding and the antenna integrated in the shielding. (d) Cut 3D-model side view of the resonator in the shielding.} 
\end{figure}

For this particular setup we have chosen a 3-loop 2-gap resonator (based on Bruker Biospin; ER 4118X-MS5).\cite{Fritsch1989,Brunner1990a,Loewenstein} The resonator was fabricated in-house from high purity silver. A 3D model of the resonator is shown in Fig.~\ref{fig4}~(a). Two large outer loops with a kidney-like shape minimize the footprint of the resonator, while the smaller inner loop is circular. Each gap connects one outer loop with the inner loop. The groove on the side of the resonator is needed to place the resonator in a fitting \textcircled{11}. The dimensions of the resonator are given in the technical drawing of the resonator shown in Fig.~\ref{fig4}~(b).

To prevent radiation losses and therefore maintain high $Q$-factors, the 3-loop 2-gap resonator has to be shielded.\cite{Froncisz1982,Wood1984} In our case, the shielding (\textcircled{13} in the explosion view of the microwave assembly in Fig.~\ref{fig3}~(b)) is a tube made of a thin copper foil, in which the resonator is placed as shown in the bottom view in Fig.~\ref{fig4}~(c).  The thickness of the copper foil is $\unit{30}{\mu m}$, which is larger than the skin depth $\delta_{\nu=\unit{9}{GHz}} \approx \unit{1}{\mu m}$ at $\unit{9}{GHz}$, but thin enough to be virtually transparent for a magnetic field modulation at typical frequencies of approximately $\unit{10}{kHz}$ with a corresponding skin depth of $\delta_{\nu=\unit{10}{kHz}} \approx \unit{1}{mm}$.\cite{Poole1997} In addition, the shielding serves as a holder for the microwave feed line \textcircled{10} and the antenna (drawn in black in Fig.~\ref{fig4}~(c)). The antenna is placed underneath one outer loop of the resonator and exhibits the same curvature as the outer loop (Fig.~\ref{fig4}~(c)) to avoid field inhomogeneities at the inner loop.\cite{Loewenstein} To better visualize the relative orientation of the resonator, the shielding, the antenna and the microwave feed line, a cut 3D-model side view of the resonator in the shielding is shown in Fig.~\ref{fig4}~(d). To mount the shielding in the microscope head, the shielding is fixed with a fitting \textcircled{15} made of the thermoplastic PEEK. The fitting \textcircled{15}, the shielding \textcircled{13}, the microwave feed line \textcircled{10} and the resonator \textcircled{12} are also shown in Fig.~\ref{fig3}~(b).

In addition, it must be possible to change the distance between the resonator and the antenna to achieve critical coupling, i.e. the microwave is coupled ideally to the resonator and no reflection occurs. For this purpose, the resonator is clipped into a resonator fitting \textcircled{11} also made of PEEK, which can be moved up and down by a coarse $z$-positioning unit \textcircled{9} (attocube systems; ANPz101) to adjust the distance between the resonator and the antenna. The coarse positioning unit itself is fixed at the upper part of the upper housing (see Fig.~\ref{fig3}~(c)). To avoid sterical hinderance with the AFM head \textcircled{4}, the two arms of the fitting \textcircled{11} are placed beside the AFM head.

The microwave assembly is connected via a circulator to a microwave generator (Rhode\&Schwarz; R\&S\textsuperscript{\textregistered}SMF100A). A microwave detection diode (Agilent Technologies; 8474C) is connected to the third port of the circulator to determine the $Q$-factor and the critical coupling of the resonator. To achieve microwave power levels of up to \unit{1}{W}, an X-band microwave amplifier (Kuhne electronic; KU~PA~093~MM) is employed.

To visualize the integration of the microwave assembly into the microscope head, a full model of the microscope head including the microwave assembly is shown in Fig.~\ref{fig3}~(c). 

\begin{figure}
	\centering
	\includegraphics[width=0.46\textwidth]{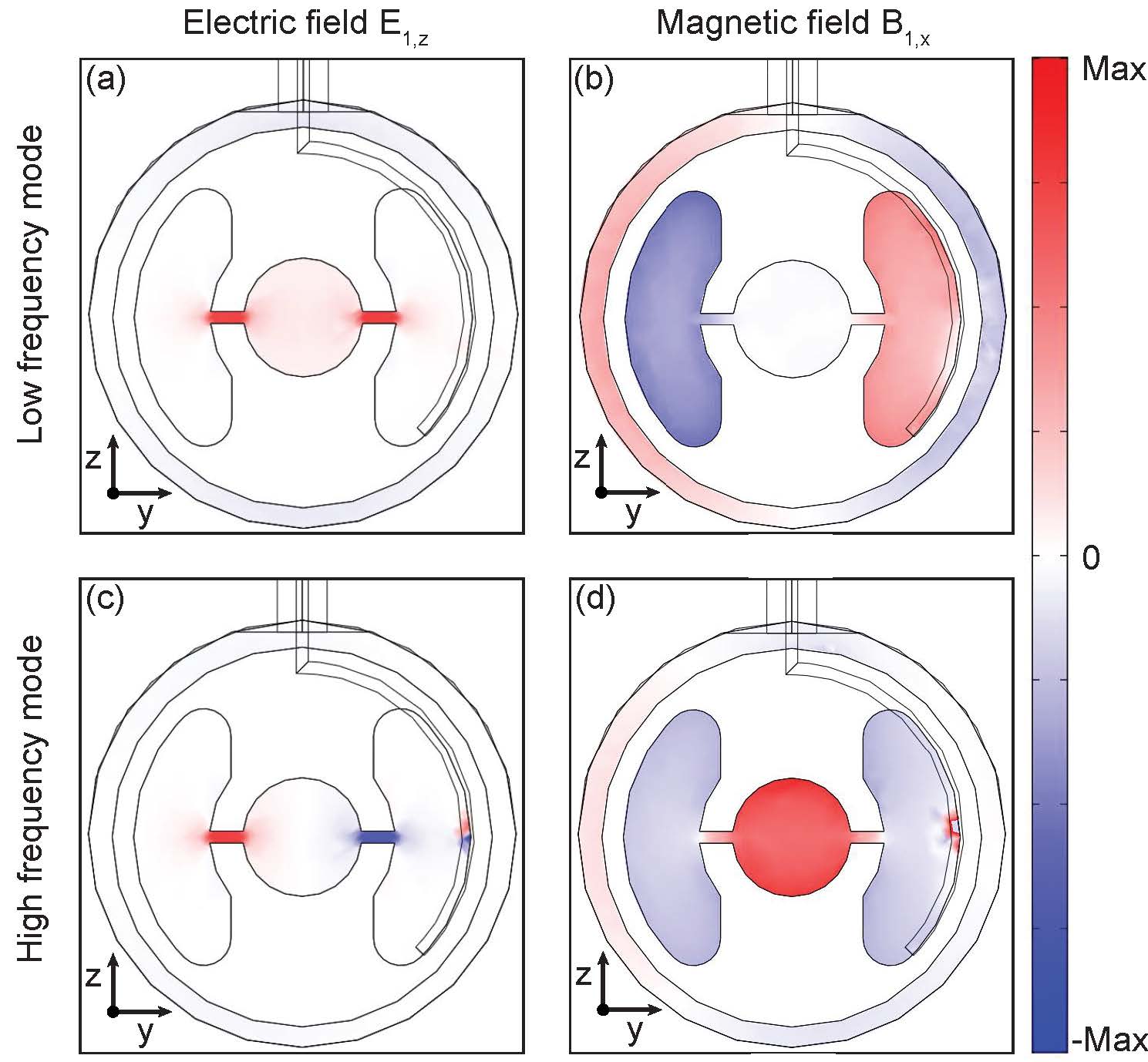}
		\caption{\label{fig5} FEM simulation of the fundamental and the high frequency mode of the 3-loop 2-gap resonator placed in a shielding. (a) $z$-component of the electric microwave field $E_1$ and (b) the $x$-component of the microwave magnetic field $B_1$ of the fundamental mode. (c) $z$-component of the microwave electric field $E_1$ and (d) the $x$-component of the microwave magnetic field $B_1$ of the high frequency mode. The results are plotted for a plane placed \unit{0.5}{mm} inside the resonator parallel to the resonator bottom surface, i.e.~the surface of the resonator facing the antenna.}
\end{figure}

For a better understanding of the electromagnetic properties of the resonator, the shielding and the antenna, we have simulated the electric field and magnetic field distributions $E_1$ and $B_1$, respectively, in the microwave assembly with a finite element method (FEM) simulation software package (COMSOL; COMSOL Multiphysics\textsuperscript{\textregistered} 3.4). For this simulation, the distance between the resonator bottom and the antenna was kept fixed at $\unit{2}{mm}$. The $z$-component of the electric field $E_1$ and the $x$-component of the magnetic field $B_1$ obtained in a plane placed \unit{0.5}{mm} inside the resonator parallel to the resonator bottom surface are plotted in Fig.~\ref{fig5}. We observe two resonant modes, the so-called fundamental or low frequency mode at $\nu_1 \approx \unit{4}{GHz}$ (Figs.~\ref{fig5}~(a)~and~\ref{fig5}~(b)) and the so-called high frequency mode at $\nu_2 \approx \unit{9}{GHz}$ (Figs.~\ref{fig5}~(c)~and~\ref{fig5}~(d)), as expected from the results published by Wood~{\it et al.}.\cite{Wood1984} For both modes, the electric field is localized between the gaps of the resonator acting as capacitors as shown in Figs.~\ref{fig5}~(a)~and~\ref{fig5}~(c).

While the differences in the $E_1$-field distribution are mostly in the relative phase within the gaps (Figs.~\ref{fig5}~(a)~and~\ref{fig5}~(c)), the differences in the magnetic field distribution $B_{1,x}$ of the two modes are more pronounced, as shown in Figs.~\ref{fig5}~(b) and \ref{fig5}~(d). For the low frequency mode (Fig.~\ref{fig5}~(b)) there is virtually no magnetic field at the inner loop of the resonator, whereas in both outer loops a strong magnetic field $B_{1,x}$ is observed. The absolute value of $B_{1,x}$ is the same in both outer loops, but the phase differs by $\pi$. This implies that the magnetic field forms a path connecting both outer loops. In case of the high frequency mode (Fig.~\ref{fig5}~(d)), a strong and homogeneous magnetic field $B_{1,x}$ is observed in the inner loop of the resonator. For both outer loops the amplitude and phase of the magnetic field $B_{1,x}$ are the same, while there is a phase difference of $\pi$ between $B_{1,x}$ in the outer loops and the inner loop, implying the existence of two closed magnetic field paths. Each path connects one outer loop with the inner loop, where the amplitudes of both paths add up to the strong $B_{1,x}$ distribution observed. For both modes disturbances of the magnetic field can only be observed at the outer loop of the resonator close to the antenna, as seen in Figs.~\ref{fig5}~(b) and \ref{fig5}~(d). This confirms the necessity to use an outer loop to couple the microwave field into the resonator to avoid disturbances of the magnetic field distribution in the inner loop at the position of the sample.

\begin{figure}
	\centering
	\includegraphics[width=0.46\textwidth]{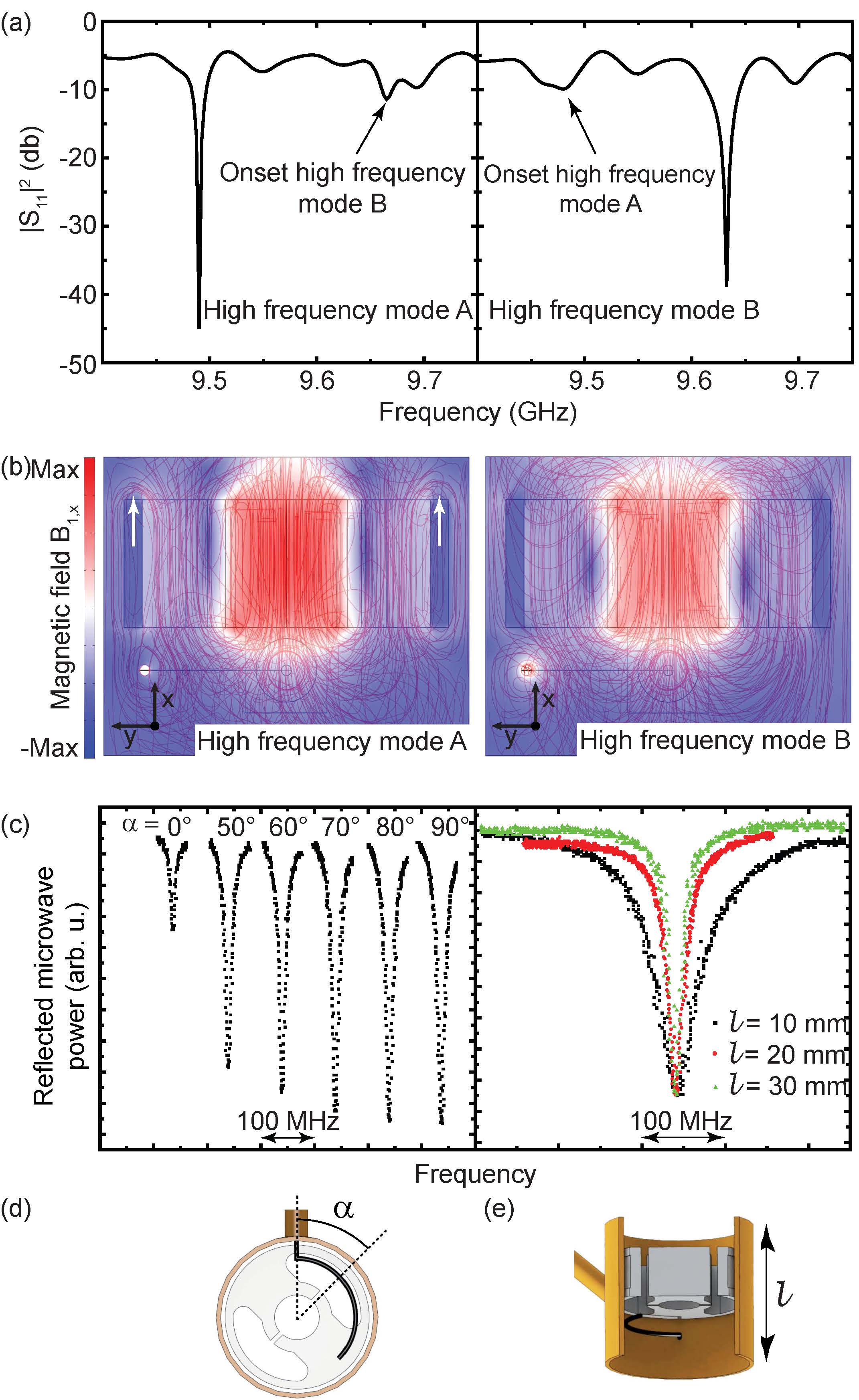}
	\caption{\label{fig6} (a) The reflection coefficient $|S_{11}|^2$ of the microwave assembly at a given resonator antenna distance (left panel) shows one high frequency mode A and the onset of a second high frequency mode B (right panel). By decreasing the resonator antenna distance the microwave can be critically coupled into the second  mode B, while only the onset of mode A is observable. (b) Magnetic field component $B_{1,x}$ and magnetic field $B_1$ field lines for modes A (left panel) and B (right panel) obtained by FEM simulations. (c)~Influence of the angle $\alpha$ between antenna and resonator on the coupling efficiency (left panel) and influence of the shielding length $l$ on the $Q$-factor of the high frequency mode A (right panel). The angle $\alpha$ and the shielding length $l$ are illustrated in (d) and (e), respectively.}
\end{figure}

To compare the results of the simulation with the actual behavior of our microwave assembly, the reflection coefficient $|S_{11}|^2$ of the empty resonator clipped into the resonator fitting and placed inside the shielding was measured.  For $\nu = 4\ldots \unit{5}{GHz}$ a sharp resonance at $\nu_1 \approx \unit{4.5}{GHz}$ was observed, which corresponds to the low frequency mode as expected (not shown). The obtained $|S_{11}|^2$-versus-frequency curves in the range of $\nu = 9.4 \ldots \unit{9.75}{GHz}$ are plotted in Fig.~\ref{fig6}~(a). The left panel of Fig.~\ref{fig6}~(a) shows a sharp resonance at $\nu_2 \approx \unit{9.5}{GHz}$, which is labeled ``high frequency mode A''. In addition, one can observe the onset of a second high frequency mode at $\nu \approx \unit{9.65}{GHz}$, which is labeled ``onset high frequency mode B'' in the left panel of Fig.~\ref{fig6}~(a). By changing the distance between the antenna and the resonator by about \unit{0.5}{mm}, it is possible to couple the resonator critically to the second high frequency mode B, while only the onset of high frequency mode A can be observed, as shown in the right hand panel of Fig.~\ref{fig6}~(a). Therefore, in contrast to the simulation with a fixed distance between the antenna and the resonator, we observe two resonant modes at $\approx \unit{9.5}{GHz}$, which are separated by $\approx \unit{150}{MHz}$ experimentally. 

By repeating our FEM simulation for various antenna resonator distances, we were able to reproduce the high frequency mode B observed experimentally. Figure~\ref{fig6}~(b) shows the magnetic field component $B_{1,x}$ (color coded) and magnetic field $B_1$ field lines (red lines) in a plane perpendicular to the resonator surface and through the middle of the gaps. For both modes the overall field distribution is the same. Two magnetic field paths connecting the inner loop with one of the outer loops are observed, as expected for the high frequency resonance of a 3-loop 2-gap resonator. Comparing the field lines for modes A and B, one can observe field lines that enclose the outer edges of the resonator for mode A (labeled with white arrows in the left panel of Fig.~\ref{fig6}~(b)), which are not observed for mode B. Despite this difference in the mode pattern, performing EDMR measurements at either mode A or B should be possible, as the magnetic field inside the inner loop is non-vanishing for both resonant modes. Nevertheless, classic EDMR experiments, i.e. using the contact geometry sketched in Fig.~\ref{fig1}~(b), performed at both modes showed that stable measurements were only always possible for mode A. Therefore, all EDMR experiments shown below were performed at resonance A.

We also have investigated how sensitively the critical coupling reacts to changes of the angle $\alpha$ between the antenna and the resonator. Here, $\alpha$ is the angle between the line defined by the center of the inner loop of the resonator and the straight part of the antenna and the line defined by the center of the inner loop of the resonator and the center of one gap of the resonator as shown in Fig.~\ref{fig6}~(d). After tuning the resonator critically to the high frequency mode A for $\alpha = 90$°, $\alpha$ is changed from $90$° to 0°. As a measure for the coupling we compare the depth of the dip observed in the reflected microwave power at the resonance frequency. The results are plotted in the left panel of Fig.~\ref{fig6}~(c). By changing $\alpha$ from $90$° to $0$°, the depth of the dip decreases. Consequently, we aligned the resonator to $\alpha = 0$° as shown in Fig.~\ref{fig4}~(c) for all experiments described in the following.  

To optimize the shielding of the resonator, we finally investigated the influence of different shielding lengths $l$ (for the definition of $l$, see Fig.~\ref{fig6}~(e)) on the high frequency mode A. The right panel of Fig.~\ref{fig6}~(c) shows the reflected microwave power versus frequency of the microwave assembly critically coupled at mode A for different shielding lengths $l$. Increasing $l$ from $10$ to $\unit{40}{mm}$ decreases the line width of the resonance significantly and, consequently, the $Q$-factor increases. Due to sterical constraints, we used a near optimum shielding length of $l=\unit{25}{mm}$ for the experiments described below.

\subsection{Modulation}

In order to apply a lock-in detection scheme, different options are available in this setup to modulate the spin-dependent current change. Microwave frequency and amplitude modulation are provided by the microwave generator. For magnetic field modulation, a pair of coils \textcircled{14} is glued onto the sides of the fitting of the microwave shielding \textcircled{15} as shown in Figs.~\ref{fig3}~(b)~and~\ref{fig3}~(c). The outer diameter of the coils is $\unit{20.5}{mm}$ and the horizontal separation $\unit{22.5}{mm}$. Although this geometry is not exactly the two-coil Helmholtz geometry,\cite{Jackson1999} which would result in a homogeneous modulation field at the position of the sample, a simulation of the geometry used reveals an inhomogeneity at the position of the sample of less than $\approx\unit{5}{\%}$ for a sample with an area of $2 \times \unit{2}{mm^2}$. The modulation coils are driven by a standard audio amplifier at typical modulation frequencies for EDMR measurements in the range of several kilohertz.\cite{Dersch1983,Lee2012}

\subsection{Sample Holder}
\label{holder}

Since most EDMR experiments are performed as measurements of the resonant spin-dependent changes of the photo-conductivity, the option to illuminate the sample has to be integrated in the setup. In addition, the need to avoid shadowing effects by the AFM cantilever requires the possibility to illuminate the sample via the substrate side. Due to the limited space in the microscope head, the position of the sample positioning stack and the fact that, in contrast to many other AFM systems, the sample is moved underneath the AFM head instead of the AFM head being moved above the sample, the illumination has to be integrated in the sample holder. The sample holder \textcircled{7} developed for this purpose is shown schematically in Fig.~\ref{fig3}~(b). It consists of a light emitting diode (LED) \textcircled{17} glued onto a base made of Delrin\textsuperscript{\textregistered}. To guide the light efficiently to the sample, a quartz glass rod \textcircled{16} is integrated in the epoxy dome of the LED, such that the rod sits $\approx \unit{1}{mm}$ above the LED dye to increase the light collection efficiency. Internal reflection guides the light to the sample on top of the rod. In addition, a thin coax cable is guided along the quartz rod toward the sample to provide one of the electrical contacts needed for cAFM and spatially resolved EDMR (not shown in Fig.~\ref{fig3}~(b)).

\subsection{Cryostat, Vibration Isolation and Magnet System}

The EDMR microscope setup is placed in a static exchange gas cryostat (Janis Research; STVP-200 for NMR) to enable measurements down to liquid helium temperature. The temperature is controlled with a cryogenic temperature controller (Cryogenic Control Systems; Model 32B). To decouple the cryostat mechanically from the surrounding setup, it is placed on an active vibration isolation unit (The Table Stable;  AVI350S/LT/LP), which is firmly attached to the electromagnet (Bruker Biospin;  B-E25). The magnet is driven by a power supply (Bruker Biospin; ER 083 CS), which is controlled by a Hall probe field controller (Bruker Biospin; BH~11~D or Lake Shore Cryotronics; 475 DSP Gaussmeter).

\subsection{Electronics and Software}

The current sensing circuit of the microscope consists of a voltage source (Yokogawa Electric Corporation; GS200 DC voltage/current source) and a  low-noise current amplifier (Femto Messtechnik; DLCPA-200). If the microscope is operated in the (p)cAFM mode, the output of the current amplifier is recorded by one of the analog-to-digital converters provided by the AFM controller (attocube systems; ASC500). For EDMR measurements, the output signal of the amplifier is fed into a dual phase lock-in amplifier (Stanford Research Systems; SR830). Depending on the type of modulation employed, the reference signal for the lock-in amplifier is either provided by the internal low frequency oscillator of the microwave source in case of microwave frequency/amplitude modulation or the internal oscillator of the lock-in amplifier is used in case of magnetic field modulation.

The operation of the AFM components is controlled by the attocube AFM electronics, consisting of an SPM controller, a piezo step controller, a piezo scan controller (attocube systems; ASC500, ANC150 and ANC200, respectively) and the control software. All other components and signals are controlled and recorded by our in-house written EDMR software package, which was adapted for this specific setup.

\section{Operation}
\label{operation}

\subsection{Demonstration of pcAFM}
\label{pcAFM}

\begin{figure}
	\centering
	\includegraphics[width=0.46\textwidth]{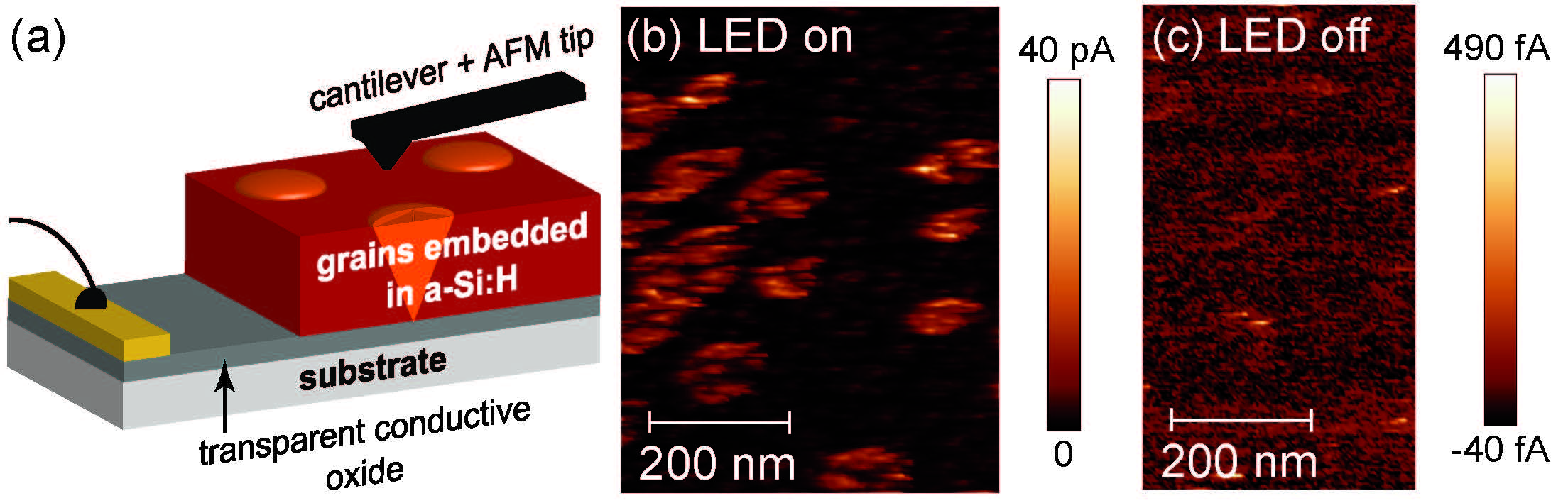}
		\caption{\label{fig7} (a) To test the absence of an influence of the AFM detection laser on the current measurements a sample with microcrystalline silicon grains embedded in amorphous silicon was used. The back contact, a transparent conductive oxide layer, was electrically connected via a bond wire and a PtIr-coated conductive AFM tip was scanned across the sample surface. (b) The local current map obtained with intentional red illumination by an LED. The amorphous matrix appears as a dark background and the microcrystalline grains can be seen as bright spots, corresponding to high current levels. (c) Local current map obtained without red illumination. The current levels observed are significantly lower than under illumination with red light. This indicates that the photo-conductivity observed under intentional illumination is not due to strong absorption of the light of the infrared detection laser used in the EDMR microscope. Since the LED in the sample holder heats up when operated, the position of the glass rod with the sample on top changes slightly due to the thermal expansion of the LED. Therefore, the maps shown in (b) and (c) do not represent the same position on the sample.} 
\end{figure}


To demonstrate pcAFM with the EDMR microscope described above we used a \unit{100}{nm} thick, intrinsic mixed phase silicon thin-film sample, i.e.~microcrystalline silicon grains embedded in an amorphous silicon matrix, with a transparent conductive oxide (TCO; here Al-doped ZnO) layer as a back contact. This back contact allows optical charge carrier excitation through the transparent glass substrate by mounting the sample onto the sample holder with an integrated red LED for intentional illumination with an average power density of $\approx \unit{40}{mW/cm^2}$ (see Section~\ref{holder}). The TCO was contacted with a bondwire and a conductive PtIr-coated contact mode AFM tip was scanned across the sample surface as shown in Fig.~\ref{fig7}~(a). A bias voltage of \unit{-2}{V} was applied to the sample.

Figure~\ref{fig7}~(b) shows a local current map of the mixed phase sample obtained under illumination with red light, Fig.~\ref{fig7}~(c) shows a local current map without this illumination. During both measurements the infrared laser for the AFM detection was operated.

Under illumination, the crystalline silicon grains can be seen as large bright spots exhibiting high current levels of up to \unit{40}{pA} in Fig.~\ref{fig7}~(b), while the local photo-current observed on the amorphous silicon matrix is $\leq \unit{4}{pA}$ and, therefore, appears as a dark background. This is in good agreement with other pcAFM results reported on this material.\cite{Ledinsky2011,Ledinsky2012} Switching off the intentional illumination results in significantly decreased current levels as shown in Fig.~\ref{fig7}~(c). The few bright spots in Fig.~\ref{fig7}~(c) again correspond to microcrystalline grains. Since the expected dark current levels for the amorphous silicon matrix should be below $\unit{10^{-2}}{fA}$ (assuming a contact area of $20 \times \unit{20}{nm^2}$ and an a-Si:H dark conductivity $< \unit{10^{-9}}{(\Omega cm)^{-1}}$),\cite{Astakhov2009} the current fluctuations of $\approx \pm \unit{50}{fA}$ outside the grains define the noise floor of the EDMR microscope operated in the (p)cAFM mode, which is similiar to that of dedicated cAFM systems.\cite{Ledinsky2011} 

In conclusion, we find that the infrared detection laser does not affect the photo-conductivity measurements of the samples to be studied with the EDMR microscope.

\subsection{EDMR Operation}
 
\begin{figure}
	\centering
	\includegraphics[width=0.46\textwidth]{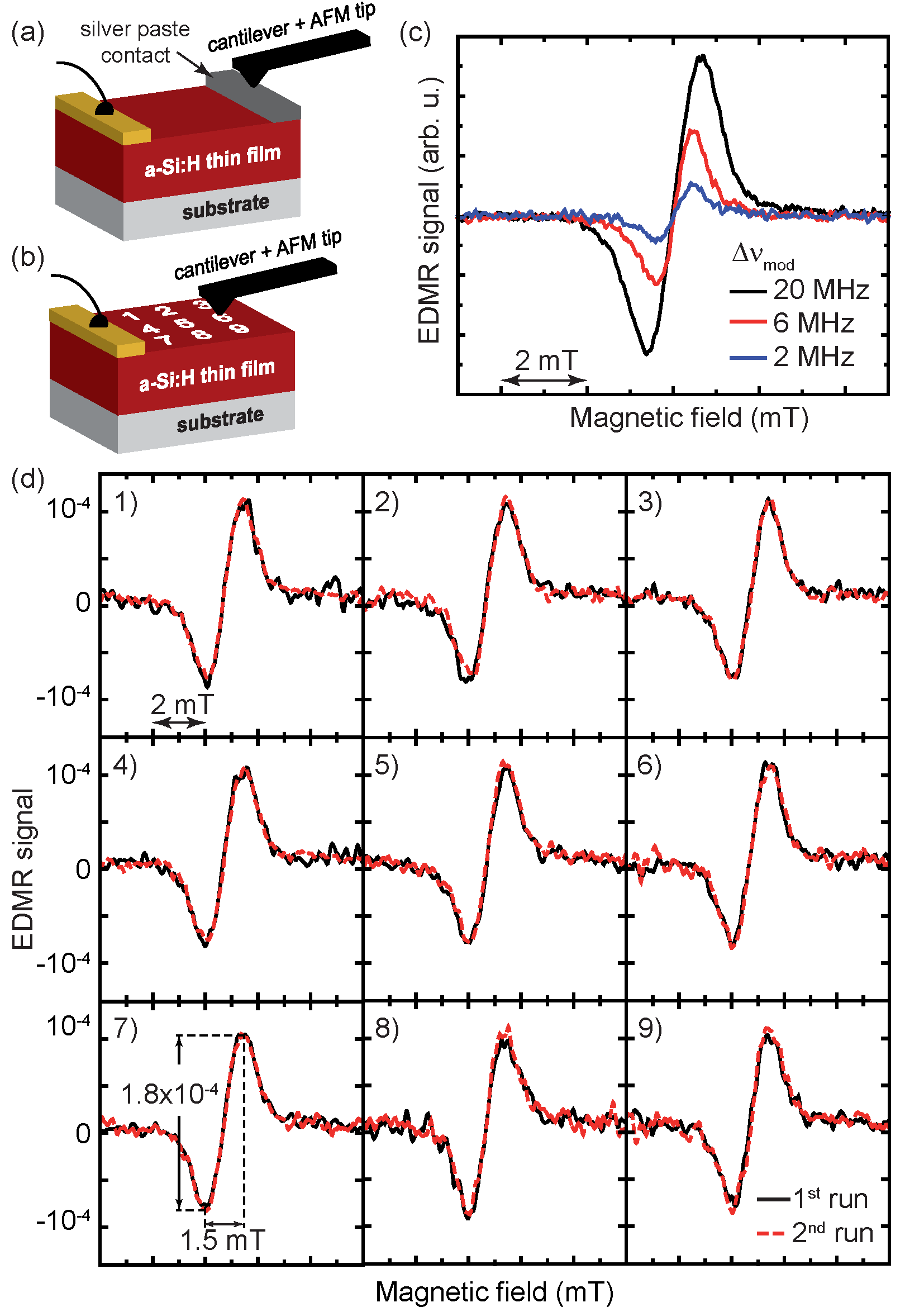}
		\caption{\label{fig8} (a) To investigate the influence of the AFM tip on the EDMR spectrum, an a-Si:H thin film sample with two coplanar contacts is used. The Cr/Au contact is wire-bonded and a conductive AFM tip is placed on the second contact made of silver paste. (b) To test the reproducibility of EDMR measurement using an AFM tip as a movable contact, an a-Si:H thin film with one evaporated Cr/Au metal contact, which is wire-bonded, is used. The AFM tip as the second contact is positioned at nine different points of a $\unit{1}{\mu m} \times \unit{1}{\mu m}$-grid on the a-Si:H film one after the other. After the first measurement run, the EDMR measurements are repeated at nominally the same positions. (c) The EDMR spectra obtained for various microwave frequency modulation amplitudes $\Delta \nu_{\mathrm{mod}}$ using the contact geometry described in (a) show only a single resonance characteristic for a-Si:H at room-temperature and no additional resonance due to the the AFM tip. (d) The EDMR spectra of the two measurement runs (black solid line = 1st run, 2nd run = red dashed line) at nine different points of the grid introduced in (b) show a good reproducibility.} 
\end{figure}

To test the EDMR capabilities and the performance of the new EDMR microscope, the feasibility of EDMR measurements with conductive AFM tips was investigated. For these experiments PtIr- and diamond-coated AFM tips with an $n$-type silicon core as well as full-metal AFM tips made of tungsten were chosen. The sample studied is a \unit{500}{nm} thick undoped a-Si:H thin film on a glass substrate with two coplanar contacts, namely one evaporated Cr/Au contact and one contact made of silver paste as shown in Fig.~\ref{fig8}~(a). As a first step the Cr/Au contact was connected via a bond wire and on the silver paste contact a conductive AFM tip was placed. The sample was again mounted on a sample holder with a red LED (see Secs.~\ref{holder} and \ref{pcAFM}).

The results of three EDMR measurements performed with an AFM tip made of tungsten for various microwave frequency modulation amplitudes $\Delta \nu_{\mathrm{mod}} = \Delta \nu_{\mathrm{mod,pp}}/2$, where $\Delta \nu_{\mathrm{mod,pp}}$ is the peak-to-peak modulation amplitude, are exemplarily shown in Fig.~\ref{fig8}~(c). Plotted are the resonant changes of the photo-current $\Delta I$ normalized to the steady state photo-current $I$. The EDMR spectrum obtained for $\Delta \nu_{\mathrm{mod}} = \unit{2}{MHz}$ (full blue line in Fig.~\ref{fig8}~(c)) consists of a single resonance at $g \approx 2.005$ with a peak-to-peak line-width $\Delta B_{\mathrm{pp}} \approx \unit{0.8}{mT}$, which is the standard EDMR signature of a-Si:H at room temperature.\cite{Dersch1983,Stutzmann2000} In addition, Fig.~\ref{fig8}~(c) also shows that the amplitude and, for large modulation amplitudes, the peak-to-peak linewidth of the observed resonance increase with increasing modulation amplitude, as expected.\cite{Poole1997} The same results were obtained for PtIr- and diamond-coated AFM tips with a crystalline silicon core (not shown). Therefore, the EDMR spectra in Fig.~\ref{fig8}~(c) show that using AFM tips as a contact does not add any spurious tip-related resonances to the EDMR spectrum at $g \approx 2$ at room temperature.

In a second step the a-Si:H film was contacted directly with a conductive AFM tip to test the reproducibility of the electrical contact between the AFM tip and the sample. For this experiment a \unit{500}{nm} thick annealed a-Si:H thin film on a glass substrate with a metal contact on one side was used as shown in Fig.~\ref{fig8}~(b). As a second contact the conductive AFM tip was placed directly onto the a-Si:H thin film. The tip was positioned at nine different points of a $\unit{1}{\mu m} \times \unit{1}{\mu m}$-grid on the a-Si:H thin film and at each point an EDMR spectrum was recorded (black line in Fig.~\ref{fig8}~(d)). After this first measurement run, the EDMR measurements were repeated at each point (red dotted line). The illumination parameters for this experiment were the same as discussed above. Frequency modulation with a modulation amplitude $\Delta \nu_{\mathrm{mod}} = \unit{20}{MHz}$ and a microwave power level of $P_{\mathrm{mw}} \approx \unit{900}{mW}$ was used to increase the signal-to-noise ratio.\cite{Poole1997,Brandt1998} 

Also this experiment was performed with various AFM tips with different conductive coatings and different force constants. A stable contact could only be obtained for contact forces $>\unit{1}{\mu N}$. In analogy to Scanning Spreading Resistance Measurements (SSRM) performed on crystalline silicon, we believe that these high contact forces are needed to penetrate the native oxide formed on the a-Si:H film surface.\cite{Wolf1996a,Wolf1998,Trenkler2000,Street1991} Due to the high contact forces only AFM tips coated with conductive diamond were durable enough to perform the experiment while metal-coated tips failed due to destroyed conductive coatings. This experimental observation agrees well with the results of other AFM, cAFM and SSRM studies.\cite{OShea1995b, Lantz1997, Lantz1998, Bhushan2008}

The most stable contact was achieved applying contact forces of $\approx \unit{4}{\mu N}$ with an AFM tip coated with conductive diamond and a force constant of $\approx \unit{43}{N/m}$. The resulting EDMR spectra are plotted in Fig.~\ref{fig8}~(d). For all nine points, the peak-to-peak EDMR signal amplitude is $\Delta I_{\mathrm{pp}}/I \approx 1.8 \times 10^{-4}$ and the peak-to-peak linewidth is $\Delta B_{\mathrm{pp}} \approx \unit{1.5}{mT}$, as expected for the intentional overmodulation used in this experiment. The spectra recorded at each position do not differ from each other within the noise level observed in this experiment. This suggests, that conductive diamond-coated AFM tips and contact forces of several $\mu\mbox{N}$ are needed to achieve a reliable contact between the AFM tip and the a-Si:H film reproducibly.

\begin{figure}
	\centering
	\includegraphics[width=0.46\textwidth]{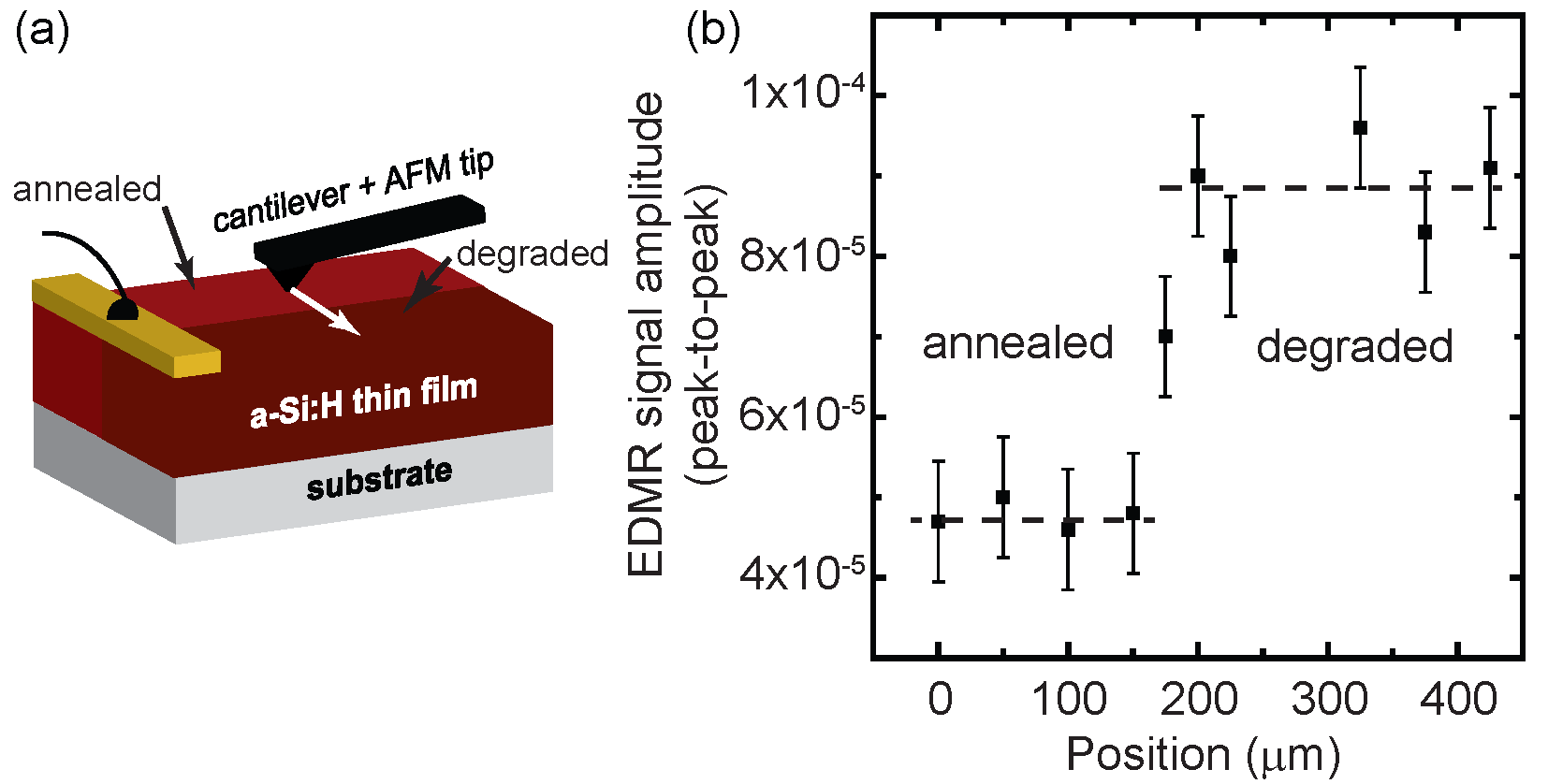}
		\caption{\label{fig9} (a) To demonstrate that the EDMR microscope is capable of detecting an EDMR contrast, an a-Si:H thin film sample was used. One half of the sample is in the annealed state, i.e.~has a low defect density, while the other half is in the degraded state, i.e.~high defect density. The AFM tip is moved parallelly to the evaporated Cr/Au metal contact. (b) The peak-to-peak EDMR signal amplitude recorded for various points along the path of the AFM tip. At \unit{175}{µm} an abrupt increase of the EDMR signal amplitude is observed, as expected for the sample structure shown in (a).} 
\end{figure}

To demonstrate, that the EDMR microscope is capable of detecting an EDMR contrast, a \unit{500}{nm} thick undoped a-Si:H thin-film sample with a varying defect density was prepared by first annealing it at $T=\unit{420}{K}$ for 1 hour. Afterwards, one half of the sample was degraded by illumination with red light (power density $\approx \unit{300}{mW/cm^{2}}$) for 18 hours,\cite{Staebler1977} while the other half of the sample was protected from the light with a shadow mask. As a result, one half of the sample is prepared in the annealed state, i.e. this half has a low defect density, and the other half of the sample is in a degraded state, i.e. that half has a high defect density. The resulting sample structure is sketched in Fig.~\ref{fig9}~(a). On one side of the sample an evaporated metal contact is placed perpendicularly to the border between the annealed and the degraded part of the sample and is connected electrically with a bond wire. A conductive diamond-coated AFM tip is moved parallelly to the metal contact at a distance $> \unit{50}{\mu m}$. This distance is necessary, since a bias of \unit{90}{V} is applied between the metal contact and the AFM tip for this experiment. If the AFM tip is closer than \unit{50}{\mu m} to the metal contact, sudden jumps of the AFM tip were observed for bias voltages $>\unit{40\ldots 50}{V}$ due to the electrostatic interaction between the conductive AFM tip and the metal contact. The illumination parameters are the same as mentioned earlier, but $\Delta \nu_{\mathrm{mod}} = \unit{10}{MHz}$ and $P_{\mathrm{mw}} = \unit{200}{mW}$ in this case, which means that the peak-to-peak signal amplitudes shown in Fig.~\ref{fig9}~(b) must be multiplied by 4 for comparison with Fig.~\ref{fig8}~(c) due to the differences in modulation amplitudes and microwave power levels.\cite{Poole1997,Brandt1998}

Figure~\ref{fig9}~(b) shows the peak-to-peak EDMR signal amplitude $\Delta I_{\mathrm{pp}}/I$ versus the position of the AFM tip. At the beginning (position \unit{0}{\mu m}) the AFM tip is placed in the annealed part of the sample. For this position an EDMR signal amplitude $\Delta I_{\mathrm{pp}}/I \approx 5 \times 10^{-5}$ is observed as expected for the experimental parameters used.\cite{Street1982} Moving the AFM tip through the annealed region, no change of the EDMR signal amplitude is observed. Starting at \unit{150}{\mu m} the EDMR signal amplitude increases until it reaches its new average value of $\Delta I_{\mathrm{pp}}/I \approx 8\ldots 9 \times 10^{-5}$ at \unit{200}{\mu m} in good agreement with the EDMR signal amplitudes reported for degraded undoped a-Si:H at room temperature.\cite{Street1982} The transition occurs on a length scale of $\approx \unit{50}{\mu m}$ as expected for a distance of $\approx \unit{50}{\mu m}$ between the tip and the evaporated metal contact.

\begin{figure}
	\centering
	\includegraphics[width=0.46\textwidth]{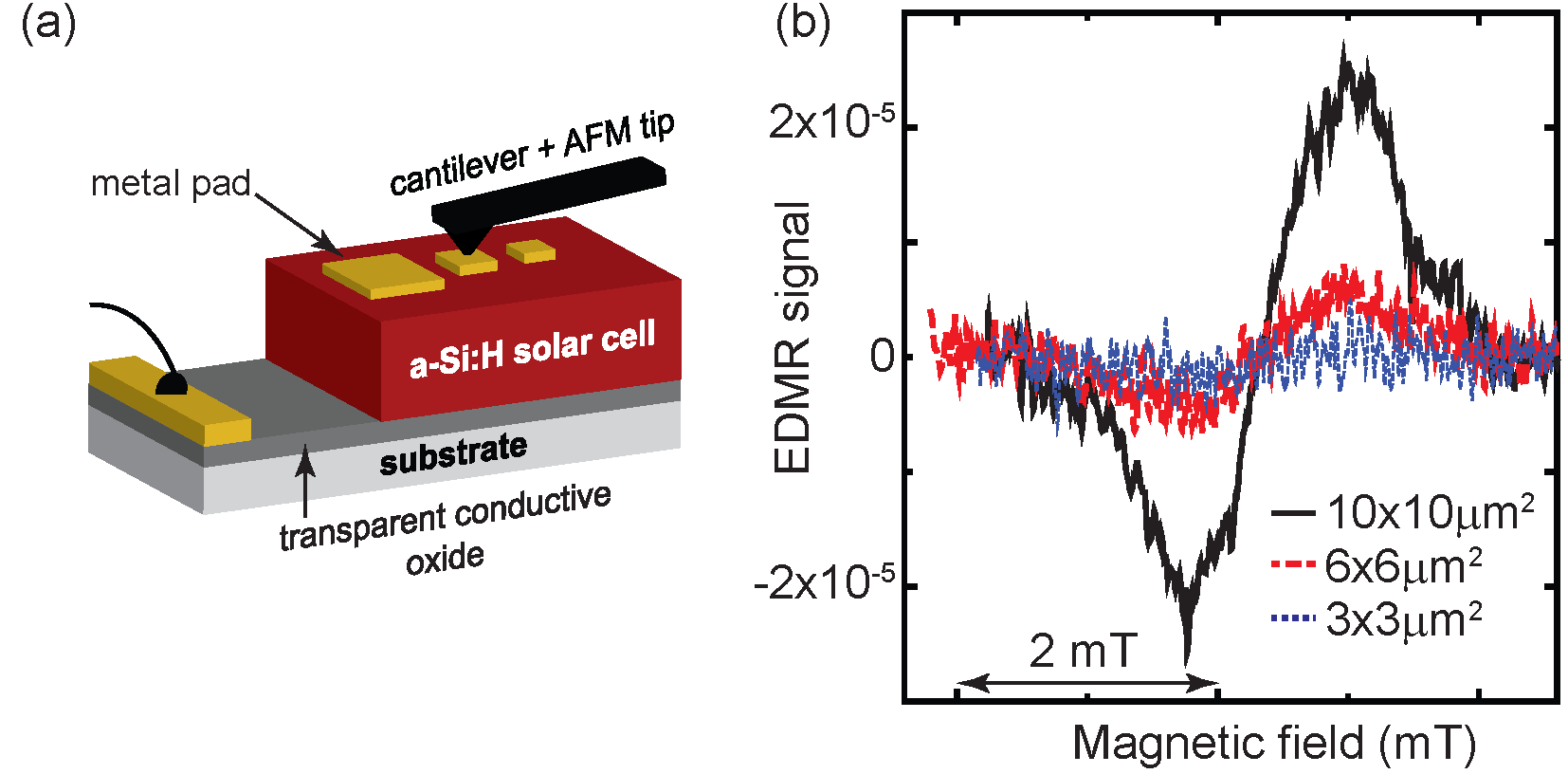}
		\caption{\label{fig10} (a) To investigate the sensitivity of the EDMR microscope quadratic Cr/Au metal contact pads were evaporated on an amorphous silicon solar cell. The pad areas range from $10 \times \unit{10}{\mu m^2}$ down to $0.1 \times \unit{0.1}{\mu m^2}$. A transparent conductive oxide (TCO) layer was used as a back contact. The individual metal pads were contacted with a conductive AFM tip to record the corresponding EDMR spectrum. (b) For three pad sizes ($10 \times \unit{10}{\mu m^2}$, $6 \times \unit{6}{\mu m^2}$ , $3 \times \unit{3}{\mu m^2}$) an EDMR signal was observed.} 
\end{figure}

After demonstrating the capability of the EDMR microscope to detect an EDMR contrast, we turn to the sensitivity of the new setup. For this experiment an a-Si:H solar cell in the degraded state is used, prepared as outlined for the thin film above. The structure of the solar cell sample is shown in Fig.~\ref{fig10}~(a). The first layer on the glass substrate is a flat \unit{700}{nm} thick TCO film followed by an a-Si:H pin-solar cell. The $p$- and the $n$-layer thickness is $\approx \unit{30}{nm}$ and the $i$-layer thickness $d_i = \unit{750}{nm}$. The upper limit of the defect density $n_{\mathrm{db}}$ in this $i$-layer is estimated to be $\approx \unit{10^{17}}{cm^{-3}}$.\cite{Lips1993,Lips1993a} Quadratic Cr/Au metal contact pads were evaporated on top of the $n$-layer. The area $A_{\mathrm{pad}}$ of these pads ranges from $10 \times \unit{10}{\mu m^2}$ down to $0.1 \times \unit{0.1}{\mu m^2}$. The TCO was electrically contacted with a bond wire and a PtIr-coated AFM tip with a force constant of $\unit{3}{N/m}$ was placed on the different metal pads as a second contact. The solar cell was biased at \unit{1.1}{V} in forward direction and illuminated with blue light ($\lambda \approx \unit{430}{nm}$, power density $\approx \unit{80}{mW/cm^2}$). Reference measurements with a macroscopic solar cell sample have shown that these parameters correspond to the highest spin-dependent current change $\Delta I$ that could be observed stably. While even higher defect densities, for example generated by e-beam irradiation, would further increase $\Delta I$,\cite{Kazanskii1986,Schneider1987} due to the long measurement times involved here, light-induced annealing as well as thermal annealing due to the thermal losses of the LED in the sample holder would lead to a strong reduction of the EDMR signal amplitude of e-beam degraded samples, so that the stable light-degraded state was used.\cite{Graeff1993,Street1982} To increase the signal-to-noise ratio frequency modulation with a modulation amplitude $\Delta \nu_{\mathrm{mod}} = \unit{20}{MHz}$ was used again.

The resulting spectra for the $10\times\unit{10}{\mu m^2}$, the $6\times\unit{6}{\mu m^2}$ and the $3\times\unit{3}{\mu m^2}$ pads are shown in Fig.~\ref{fig10}~(b). For the $10\times\unit{10}{\mu m^2}$ pad an EDMR signal amplitude $\Delta I_{\mathrm{pp}}/I$ of $4 \times 10^{-5}$ is observed after averaging 159 magnetic field sweeps. Decreasing the pad area down to $6\times\unit{6}{\mu m^2}$ and $3\times\unit{3}{\mu m^2}$ reduces the EDMR signal amplitude to $1 \times 10^{-5}$ (114 sweeps) and $\approx 4 \times 10^{-6}$ (180 sweeps), respectively.  For all metal pads $<3\times\unit{3}{\mu m^2}$ no EDMR signal could be observed within 200 magnetic field sweeps. The detected resonant current changes $\Delta I_{\mathrm{pp}}$ for the different pad areas are $\Delta I_{10 \times \unit{10}{\mu m^2}} \approx \unit{680}{fA}$, $\Delta I_{6 \times \unit{6}{\mu m^2}} \approx \unit{90}{fA}$ and $\Delta I_{3 \times \unit{3}{\mu m^2}} \approx \unit{20}{fA}$. The EDMR microscope developed here, therefore, reaches the same high sensitivity in terms of $\Delta I$, that is found for state-of-the art EDMR spectrometers using fixed contacts only. 

It is interesting to note that the observed peak-to-peak EDMR signal amplitude $\Delta I_{\mathrm{pp}}/I$ is decreasing with decreasing pad area, while a constant $\Delta I_{\rm pp}/I$ would be expected in the case of a linear scaling of $\Delta I_{\mathrm{pp}}$ and $I$ with $A_{\rm pad}$. One possible reason could be electrical field enhancement at the edges of the rectangular metal pads, which could change the transport properties locally resulting in a decrease of the observed signal amplitude.\cite{Nardi2012} 
%

\begingroup
\squeezetable
\begin{table*}
\caption{\label{tablesens1} Summary of the sensitivity chararcterization of the EDMR microscope shown in Fig.~\ref{fig10}.}
\begin{ruledtabular}
\begin{tabular}{ccccccccccc}
Pad size & $\Delta I_{\mathrm{pp}}/I$ & $I$ (nA) & $\Delta I_{\mathrm{pp}}$ (fA) & Number of & SNR & SNR\textbar$_1$\footnotemark[1] & Number of & Sensitivity $s$ & Shot noise & Overall noise (fA) \\
$({\rm \mu m^2})$ &  &  &  & sweeps &  &  & spins $N$\footnotemark[2] & ($\mbox{spins}/\sqrt{\mbox{Hz}})$\footnotemark[3] & $I_{\rm shot}$ (fA) & \\ \hline 
$10\times 10$ & $4 \times 10^{-5}$ & $\approx 17$ & $\approx 680$ & 159 & 11  & $\approx 0.87$ & $7.5 \times 10^{6}$ & $8.6 \times 10^{6}$ & $\approx 75$ & $\approx 780$ \\ \hline
$6\times 6$   & $1 \times 10^{-5}$ & $\approx 8$ & $\approx 90$  & 114 & 3.6 & $\approx 0.33$ & $2.7 \times 10^{6}$ & $8.2 \times 10^{6}$ & $\approx 50$ & $\approx 260$ \\ \hline
$3\times 3$   & $4 \times 10^{-6}$ & $\approx 4.5$ & $\approx 20$  & 180 & $\approx 1.7$   & $\approx 0.12$ & $6.7 \times 10^{5}$ & $5.6 \times 10^{6}$ & $\approx 40$ & $\approx 150$ \\ 
\end{tabular}
\end{ruledtabular}
\footnotetext[1]{SNR\textbar$_1$ = SNR/$\sqrt{\mbox{number of sweeps}}$.}
\footnotetext[2]{Calculated according to Eq.~\eqref{numberofspins}.}
\footnotetext[3]{Calculated according to Eq.~\eqref{sense}.}
\end{table*}
\endgroup

To evaluate the number of spins $N$ detected by the EDMR measurements presented in Fig.~\ref{fig10}, the active volume, i.e., the volume through which the current is driven is estimated. To do so, we assume that the current spreading in the $n$-layer of the solar cell can be neglected. As a consequence the active area is the pad area $A_{\rm pad}$. In addition, the EDMR signal amplitude observed for highly doped (doping concentration $\approx 1\%$) $p$- and $n$-type thin films made of a-Si:H is approximately a factor 100 less than for intrinsic films.\cite{Solomon1977} As a consequence the contribution of the $n$- and $p$-type layer of the solar cell to the observed EDMR signal amplitude can be neglected. Therefore, the number of spins $N$ is estimated as
\begin{equation}
\label{numberofspins}
N = A_{\mathrm{pad}} ~ d_i ~ n_{\mathrm{db}}, 
\end{equation}
with $d_i = \unit{750}{nm}$ being the thickness of the $i$-layer and $n_{\mathrm{db}} = \unit{10^{17}}{cm^{-3}}$ the defect density in the $i$-layer. Evaluating Eq.~(\ref{numberofspins}) for different pad areas, the estimated number of spins contributing to the EDMR signal observed experimentally ranges from $7.5\times 10^6$ spins for the $10 \times \unit{10}{\mu m^2}$-pad down to $6.7\times 10^5$ spins for the $3 \times \unit{3}{\mu m^2}$-pad as summarized in Tab.~\ref{tablesens1}. 

\section{Discussion}
\label{discussion}

To further elucidate the sensitivity of the EDMR microscope presented, we discuss the different contributions to the noise. The unavoidable shot noise contribution to the experimentally observed noise is $I_{\mathrm{shot}} = \sqrt{2 q I \Delta f}$ with $q$ being the elementary charge and $\Delta f = \unit{1}{Hz}$ the detection bandwidth of the current detection setup.\cite{Schottky1918} The shot noise level of the EDMR spectra of the different pads (c.f. Table~\ref{tablesens1}) ranges from $I_{\mathrm{shot},10 \times \unit{10}{\mu m^2}}\approx \unit{75}{fA}$ for the $10\times \unit{10}{\mu m^2}$ pad down to $I_{\mathrm{shot},3 \times \unit{3}{\mu m^2}}\approx \unit{40}{fA}$ for the $3\times \unit{3}{\mu m^2}$ pad. Comparing these levels with the microwave-induced resonant current change $\Delta I_{\rm pp}$, one can immediately see that for the $10\times \unit{10}{\mu m^2}$ pad $I_{\mathrm{shot},10 \times \unit{10}{\mu m^2}} \approx \Delta I_{10 \times \unit{10}{\mu m^2}}/10$ and for the $6\times \unit{6}{\mu m^2}$ pad $I_{\mathrm{shot},6 \times \unit{6}{\mu m^2}} \approx \Delta I_{6 \times \unit{6}{\mu m^2}}/2$, i.e.~the shot noise contribution is (significantly) less than the resonant current change $\Delta I_{\mathrm{pp}}$. For the $3\times \unit{3}{\mu m^2}$ pad $I_{\mathrm{shot},3 \times \unit{3}{\mu m^2}} = \unit{40}{fA}$ is already twice the resonant current change $\Delta I_{3 \times \unit{3}{\mu m^2}}$.

In addition to shot noise, the noise contribution of the current amplifier of \unit{43}{fA} has to be taken into account. For the EDMR measurement performed for the $3 \times \unit{3}{\mu m^2}$ pad also this contribution is higher than the resonant current change $\Delta I_{3 \times \unit{3}{\mu m^2}} \approx \unit{20}{fA}$. A higher amplification would decrease the noise contribution of the amplifier, which would however simultaneously reduce the bandwidth to less than the modulation frequency used in these experiments.

Comparing the overall noise level of the experiments, i.e. the noise level observed off resonance for the EDMR spectra in Fig.~\ref{fig10},  with the shot noise and the amplifier noise contributions clearly indicates that both contributions are important as the pad size gets smaller, but do not dominate. We believe that the major contribution to the overall noise level originates from the electrical contact between the conductive AFM tip and the metal pads. Only optimized contact forces in the range of $200\ldots \unit{500}{nN}$ resulted in stable contacts between tip and metal pad for up to a few hours, which is in good agreement with results reported by other groups.\cite{Bietsch2000, Guo2006} Even higher forces either destroyed the metal pad or the conductive coating of the AFM tip.

Consequently, the stability of the AFM tip sample contact has to be further improved to achieve an even higher current sensitivity of the EDMR microscope. Nevertheless, resonant current changes of the order of \unit{10}{fA} are detected, making this technique a highly sensitive cAFM derivative. Furthermore, increasing $\Delta I_{\mathrm{pp}} / I$ by increasing the defect density or by observing charge transport processes with an even stronger Pauli dependence up to a full Pauli blockade would lead to even higher sensitivities of the microscope developed.

To compare the performance of the EDMR microscope with the published performance of other ESR/EDMR spectrometers, we determined the sensitivity of the EDMR microscope. The sensitivity $s$ is defined as the minimum number of spins which can be detected at a bandwidth of $\unit{1}{Hz}$ and a signal-to-noise ratio (SNR) $=1$ and is calculated as\cite{Boero2003} 
\begin{equation}
\label{sense}
s = \frac{1}{\mbox{SNR}}~\frac{N}{\sqrt{\Delta f}},
\end{equation}
with $\Delta f$ being the experimental detection bandwidth. Applying Eq.~(\ref{sense}) to the data obtained for the sensitivity measurements of the EDMR microscope (Tab.~\ref{tablesens1}) yields an average sensitivity $s \approx 8 \times \unit{10^6}{spins/\sqrt{Hz}}$. 

To put this into context, sensitivities reported by various groups for cwESR/cwEDMR setups as well as pulsed ESR setups are summarized in Tab.~\ref{tablesens3}. Due to the different scalings of sensitivity with the linewidth, we explicitly state the linewidth of the corresponding resonance as well as the material system and the operating temperature. Using the definition Eq.~\eqref{sense} of the sensitivity Maier realizes a sensitivity of $\approx \unit{6\times 10^9}{spins/\sqrt{Hz}}$ using a Bruker ELEXYS X-band ESR-spectrometer and an ER4122-SHQ super-high-Q resonator at room temperature.\cite{Maier1997a} Using planar microresonators with resonant frequencies $\approx 13 \ldots \unit{14}{GHz}$ Narkowicz~\textit{et al.} achieve sensitivities up to $\unit{4.3\times 10^9}{spins/\sqrt{Hz}}$ at room temperature.\cite{Narkowicz2005,Narkowicz2008} 

Using a $\unit{5}{\mu m}$ surface loop gap resonator (resonance frequency $\unit{15\ldots 17}{GHz}$) for pulsed ESR imaging and $\gamma$-irradiated SiO$_2$ as a sample Blank~\textit{et al.}~were able to achieve a sensitivity of $\unit{3\times 10^7}{spins/\sqrt{Hz}}$ at room temperature.\cite{Twig2013} For a $\unit{20}{\mu m}$ surface loop gap resonator and a phosphorus-doped isotopically pure crystalline $^{28}$Si sample, they report a sensitivity of $\unit{3.2\times 10^4}{spins/\sqrt{Hz}}$ at $\unit{10}{K}$.\cite{Blank2013}

In contrast to electron spin resonance there are only very few publications discussing the sensitivity of EDMR experiments. To our knowledge Kawachi~\textit{et al.}~are the only ones reporting on the sensitivity of a room-temperature X-band cwEDMR experiment. Investigating a-Si:H thin-film transistors they achieved a sensitivity of $\approx \unit{4.5 \times 10^5}{spins/\sqrt{Hz}}$, however using $B_1$ fields significantly higher than those available here.\cite{Kawachi1996,Kawachi1997} McCamey~\textit{et al.}~performed a systematic study on the sensitivity achieved with X-band EDMR at liquid helium temperature using phosphorus doped crystalline silicon samples. They were able to detect as little as $\approx 100$ phosphorus donor electron spins corresponding to a sensitivity of $\approx \unit{400}{spins/\sqrt{Hz}}$.\cite{McCamey2006a,Huebl2007}

Comparing the room temperature sensitivity $s \approx 8 \times \unit{10^6}{spins/\sqrt{Hz}}$ achieved with the EDMR microscope with the sensitivities reported by other groups for various ESR and EDMR experiments at room temperature indicates that the EDMR microscope is competitive. As already pointed out, we believe that the noise  originating from the tip-metal pad contact appears to be the limiting factor. As drift and vibrational noise of the system and in particular of the AFM cantilever will be reduced at lower temperatures, an even higher sensitivity can be expected for low temperature operation of the EDMR microscope. EDMR experiments with phosphorus-doped crystalline silicon samples have shown, that EDMR measurements with the EDMR microscope using a conductive AFM tip as one contact  are possible at liquid helium temperature. Nevertheless, the transport mechanisms giving rise to the EDMR signal observed and the properties of the tip-sample contact at cryogenic temperatures are still under investigation.


\begingroup
\squeezetable
\begin{table}
\caption{\label{tablesens3} Summary of the sensitivities of ESR/EDMR experiments reported.}
\begin{ruledtabular}
\begin{tabular}{lcccc}
Method & Temp. & Sensitivity $s$ & Material / & Ref.  \\ 
& & ($\mbox{spins} / \sqrt{\mbox{Hz}}$) & peak-to-peak & \\
& &  & linewidth & \\ \hline
EDMR	& RT & $\approx 8 \times 10^{6}$ & a-Si:H & This work\\
microscope &  & & \unit{0.8}{mT} & \\ \hline
cwESR\footnotemark[1]		& RT & $\approx 6.6 \times 10^9$ & Tempol in benzene & \citenum{Maier1997a}\\
& & & \unit{0.3}{mT} & \\ \hline
cwESR		& RT & $4.3 \times 10^{9}$ & DPPH & \citenum{Narkowicz2005}\\
& & & \unit{0.14}{mT} & \\ \hline
cwEDMR\footnotemark[2]	& RT & $\approx 4.5 \times 10^5$ & a-Si:H & \citenum{Kawachi1996}\\
& & & \unit{0.8}{mT} & \\ \hline
cwEDMR\footnotemark[3]	& \unit{5}{K} & $\approx 400$ & c-$^{\mathrm{nat}}$Si:P & \citenum{McCamey2006a,Abe2010}\\
& & & \unit{0.22}{mT} & \\ \hline
pESR		& RT & $3 \times 10^{7}$ & $\gamma$-irradiated $\mbox{SiO}_2$ & \citenum{Blank2013,Weeks1956}\\
& & & $\leq \unit{0.1}{mT}$ & \\ \hline
pESR		& \unit{10}{K} & $3.2 \times 10^{4}$ & c-$^{\mathrm{28}}$Si:P & \citenum{Blank2013,Tyryshkin2003}\\
& & & $\approx \unit{0.008}{mT}$ & \\
\end{tabular}
\end{ruledtabular}
\footnotetext[1]{Maier reports an absolute sensitivity limit of $\unit{9 \times 10^8}{spins/G}$ normalized for a SNR = 1. To obtain the sensitivity in $\mbox{spins}/\sqrt{\mbox{Hz}}$ we multiply this value by the square of the linewidth $(\unit{3}{G})^2$. The detection bandwidth is estimated to be $\Delta f = 1/\mbox{conversion time} = 1/\unit{0.66}{s} \approx \unit{1.52}{Hz}$.} 
\footnotetext[2]{Spins observed $\approx 10^5$, single sweep SNR$\approx 0.44$, detection bandwidth $\Delta f \approx \unit{0.25}{Hz}$.}
\footnotetext[3]{Spins observed $\approx 100$, single sweep SNR=0.5, detection bandwidth $\Delta f \approx \unit{0.5}{Hz}$.} 
\end{table}
\endgroup

\section{Conclusion}

We presented the design, the successful implementation and the operation of a combined (p)cAFM  and EDMR setup by integrating the microwave assembly needed for EDMR  consisting of a 3-loop 2-gap resonator, the shielding and the microwave feed line into a commercially available AFM, a concept which could be used for microwave delivery in ODMR-, MRFM- or scanning thermal microscopy-based setups for microwave spectroscopy.\cite{Albrecht2013,Bennett2013,Meckenstock2008} This EDMR microscope is capable of operating between room temperature and liquid helium temperature and under magnetic fields of up to several Tesla. We demonstrated that it is possible to use a conductive AFM tip as a movable contact for EDMR measurements realizing a highly sensitive cAFM derivative capable of detecting resonant current changes of the order of \unit{10}{fA}. Using this contact configuration we were able to spatially resolve the EDMR contrast expected in a partially degraded amorphous silicon sample. The spin sensitivity of the EDMR microscope achieved is $\approx 8\times \unit{10^6}{spins/\sqrt{Hz}}$ at room temperature. The main limitation in terms of spin sensitivity appears to be the noise originating from the conductive AFM tip-sample contact. This microscope, therefore, allows to measure EDMR with a movable contact at the same high sensitivity as conventional EDMR spectrometers using two fixed, evaporated contacts, and to add spectral information to spatially resolved conductivity measurements for an in-depth understanding of electrical transport phenomena in micro- and nanostructures constituting an sEDMR setup as defined in Sec.~\ref{introduction}. 

\begin{acknowledgments}
We gratefully acknowledge helpful discussions with K. Karrai, C. Mitzkus, C. Bödefeld and R. Pohlner (attocube systems, Germany). We also would like to thank J. Löwenstein (FU Berlin) for supplying a brass version of the resonator and A. Kupijai for help with model rendering. This work was financially supported by  Bundesministerium für Bildung und Forschung (Grant No. 03SF0328B, EPR-Solar) and Deutsche Forschungsgemeinschaft (Grant No. SFB 631, C3). 
\end{acknowledgments}

\bibliographystyle{aipnum4-1}

%

\end{document}